\documentclass[ps,prb,twocolumn,showpacs,superscriptaddress]{revtex4-2}  
\usepackage{amsmath}
\usepackage{amssymb}
\usepackage{graphicx}
\usepackage{suffix}
\usepackage{mathtools}
\usepackage{tabularx}
\usepackage{gensymb}
\usepackage{placeins}
\usepackage{float}
\usepackage{braket}
\usepackage{array}
\newcolumntype{Z}{>{\centering\let\newline\\\arraybackslash\hspace{0pt}}X}

\begin{document}
\title{Spin interactions and topological magnonics in chromium trihalide CrClBrI}

\author{Eliot Heinrich}
\address{Department of Physics, Boston College, 140 Commonwealth Avenue, Chestnut Hill, Massachusetts 02467, USA}
\author{Xin Li}
\address{Department of Physics, Boston College, 140 Commonwealth Avenue, Chestnut Hill, Massachusetts 02467, USA}
\author{Benedetta Flebus}
\address{Department of Physics, Boston College, 140 Commonwealth Avenue, Chestnut Hill, Massachusetts 02467, USA}

\begin{abstract}

The discovery of spontaneous magnetism in van der Waal (vdW) magnetic monolayers has opened up an unprecedented platform for investigating magnetism in purely two-dimensional systems. Recently, it has been shown that the magnetic properties of vdW magnets can be easily tuned  by adjusting the relative composition of halides.
Motivated by these experimental advances, here we derive a model for a trihalide CrClBrI monolayer  from symmetry principles and we find that, in contrast to its single-halide counterparts, it can display  highly anisotropic nearest- and next-to-nearest neighbor Dzyaloshinskii-Moriya  and Heisenberg interactions. Depending on the parameters, the DM interactions are responsible for the formation of exotic chiral spin states, such as skyrmions and spin cycloids, as shown by our Monte Carlo simulations. Focusing on a ground state  with a two-sublattice  unit cell,  we find spin-wave bands with nonvanishing Chern numbers. The resulting magnon edge states yield a  magnon thermal Hall conductivity that changes sign as function of temperature and  magnetic field, suggesting chromium trihalides as a candidate for testing topological magnon transport in two-dimensional noncollinear spin systems.
\end{abstract}

\maketitle

\section{Introduction} While spin phenomena in two dimensions have been subjected to intense  scrutiny for decades, only recently have vdW magnets emerged as a concrete platform for the exploration of  two-dimensional (2$d$) magnetism~\cite{burch2018,gong2017,huang2017,park2016}. In most of these compounds, a long-range order is stabilized by an in-plane or out-of-plane magnetic anisotropy that circumvents the restrictions of the Mermin-Wagner theorem~\cite{mermin1966,hohenberg1967,huang2017,lee2016,wang2016,gong2017,bonilla2018,ohara2018}.
Monolayers of  chromium halides $\text{CrX}_{3}$ (X=Cl,Br,I)  have been proposed as testbed for the  Berezinskii-Kosterlitz-Thouless universality class that has been long sought in magnetic systems~\cite{berezinskii1971,kosterlitz1973,kosterlitz1974,kim2021,troncoso2020}. 
Their honeycomb lattice structure  has opened up opportunities to investigate Dirac bosons, whose statistics and interactions drastically differed from their far more scrutinized electronic counterpart~\cite{pershoguba2018}.  With  strong spin-orbit coupling (SOC) and an edge-sharing octahedra structure,  vdW ferromagnets can display a  bond-directional anisotropic exchange interaction, i.e.  the Kitaev interaction~\cite{kitaev2006,xu2018,lee2020}, providing a route for the investigation of spin liquid states with spin  $S=3/2$~\cite{xu2020:kitaev}. 
Furthermore, the lattice structure symmetry allows for next-to-nearest neighbor (NNN) out-of-plane Dzyaloshinskii-Moriya (DM) interactions. NNN DM  interactions on a honeycomb ferromagnetic lattice play a role analogous to SOC in graphene: magnons  accumulate an additional phase  upon propagation between NNN sites and topologically nontrivial edge states can emerge~\cite{chen2018,kim2016,ruckriegel2018}.

The variety of magnetic regimes  displayed by vdW magnets can be further enriched by tuning their properties through electric fields,  proximity effects or chemical doping~\cite{wang2018, behera2019, lu2017,liu2018,zhong2017,hellman2017,abramchuk2018,kondo2020d}. Recently, Tartaglia \textit{et al.}~\cite{tartaglia2020} have shown that the magnetic anisotropy of chromium halides can be continuously tuned by adjusting the relative composition of halides.  Importantly, varying the ratio of ligands not only affects the overall anisotropy, but also leads to a crystalline structure with a lower symmetry group than its stochiometric counterpart.

\begin{figure}[b!]
	\centering
	\includegraphics[width=0.5 \textwidth]{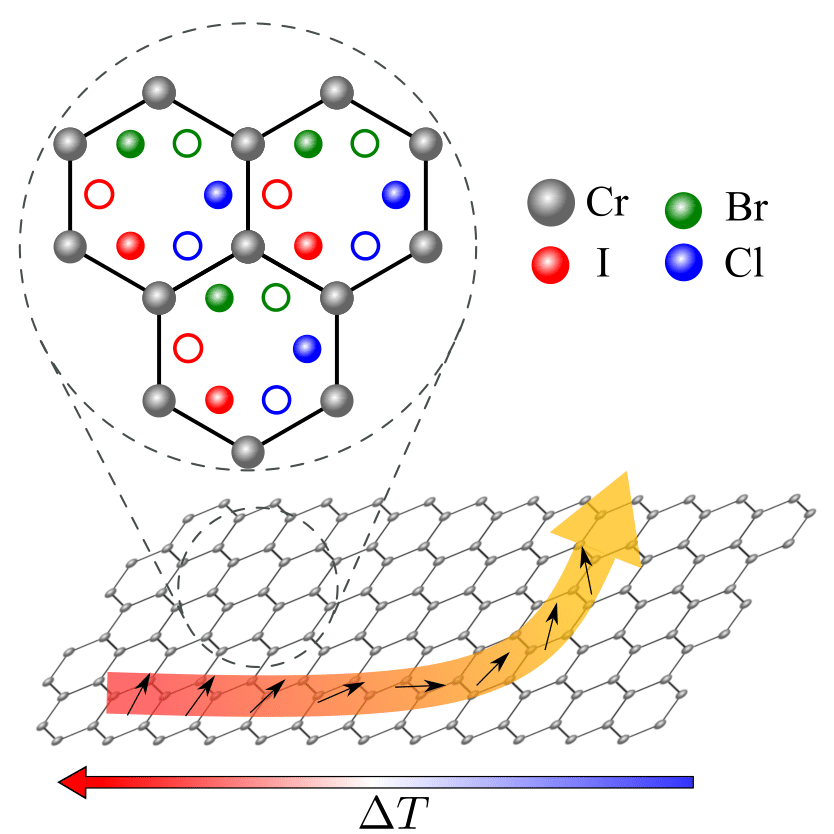}
	\caption{Lattice structure of a chromium trihalide monolayer. The magnetic atoms (Cr) are arranged in a honeycomb lattice. The Cr-Cr coupling is mediated by I, Cl and Br ligands. Solid colored dots refer to atoms above the Cr plane and open dots refer to atoms below the Cr plane. In this work, we explore the emergence of topologically protected magnon edge states that yield  a thermal Hall flow,   transverse with respect to the direction of an applied temperature gradient $\Delta T$.}
	\label{fig:crystal}
\end{figure}

Motivated by these experimental advances, in this work we investigate the magnetic properties of a chromium trihalide CrClBrI layer, shown in Fig.~\ref{fig:crystal}. We show that the richness of spin-spin interactions can lead, depending on the parameters, to topological magnon phases and to a wide array of noncollinear spin states and magnetic defects.

This work is organized as follows:
 In section II, we establish a Hamiltonian spin model for a chromium trihalide CrClBrI layer.  In section III, we explore a set of system parameters corresponding to a two-sublattice ground state. In this regime, we show that the spin-wave bands can have nonvanishing Chern number, which signals the presence of  topologically protected edge states.  We investigate the contribution of these edge states to the magnon thermal Hall effect~\cite{katsura2010,matsumoto2011,murakami2017,onose2010}. Finally, in section IV, we demonstrate using Monte Carlo techniques that our model can support  exotic noncollinear ground states such as spin cycloids and Bloch and N\'{e}el skyrmions.

\section{Model} 

\begin{figure}
	\centering
	\includegraphics[width=250px]{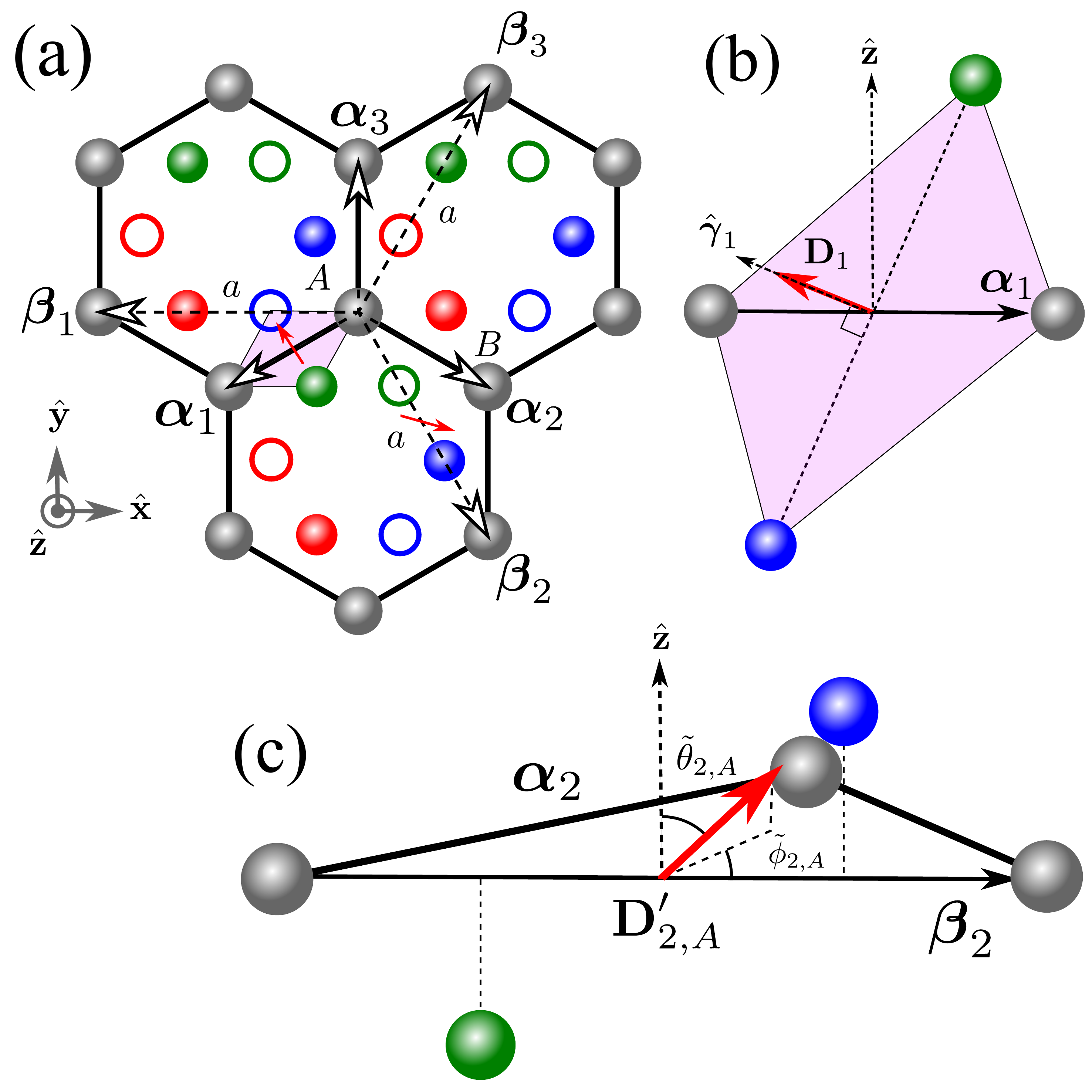}
	\caption{
	(a) The bond geometry is shown. $A$ and $B$ label the two magnetic sublattices of the honeycomb lattice, while $\boldsymbol\alpha_n$  and $\boldsymbol\beta_n$ label, respectively, the NN and NNN bond vectors, with $n=1,2,3$. The length of the NNN bond is $a$, i.e. $| \boldsymbol \beta_n | = a$. On the $\boldsymbol \alpha_1$ bond, the mirror plane is shown in purple, along with the NN DM vector (red arrow). The NNN DM vector is shown in red on the $\boldsymbol \beta_2$ vector.
	(b) The NN bond geometry along the hopping direction defined by $\boldsymbol\alpha_1$, mediated by a Cl below the plane and Br above the plane. The purple plane containing both Cr atoms and the two mediating halides is a mirror plane: by Moriya's rules, the DM vector $\textbf{D}_{1}$ (red arrow) is constrained to be perpendicular to this plane.
	(c) The NNN bond geometry along the hopping direction defined by $\boldsymbol\beta_2$. The red arrow represents the NNN DM vector $\textbf{D}'_{2,A}$. }
	\label{fig:structure}
\end{figure}

Let us consider a monolayer of chromium trihalide CrClBrI. The magnetic Cr atoms are arranged on a honeycomb lattice and each $i$th site $\mathbf{r}_{i}$ carries a spin moment $\textbf{S}_i = (S_i^x, S_i^y, S_i^z)$.  The spin-spin interactions between Cr atoms are mediated by the nonmagnetic ligands (Cl, I, and Br) lying out of the Cr plane, as shown in Fig.~\ref{fig:structure}(a). The distribution of ligands breaks the $C_3$ symmetry of the honeycomb lattice and allows interactions to be bond-dependent. The nearest-neighbor (NN) Heisenberg exchange term can be generally written as

\begin{equation} \label{jterm}
\mathcal{H}_{J}^{NN} = -\sum\limits_{\langle i,j \rangle} J_{ij} \textbf{S}_i \cdot \textbf{S}_j\,,
\end{equation}
where $\langle .. \rangle$ denotes summation over the nearest neighbors and $J_{ij}$ is the bond-dependent ferromagnetic exchange coupling. Here, $J_{ij}$  takes the values $J_1$, $J_2$, or $J_3$ for the NN bond along $\boldsymbol \alpha_1$, $\boldsymbol \alpha_2$ and $\boldsymbol \alpha_3$, respectively.
The bond geometry is shown in Fig.~\ref{fig:structure}(a).

In addition, the SOC allows for an antisymmetric exchange, i.e. a Dzyaloshinskii-Moriya (DM) interaction between both NN and NNN atoms. The NN DM interaction contribution to the Hamiltonian reads 

\begin{equation} \label{dterm}
\mathcal{H}_{DM}^{NN} = -\sum\limits_{\langle i,j \rangle} \textbf{D}_{ij} \cdot (\textbf{S}_i \times \textbf{S}_j).
\end{equation}

The DM vectors are determined by Moriya's rules \cite{moriya1960} according to the local symmetry of the bond. Similar to the NN Heisenberg interaction~\eqref{jterm}, the DM strength is bond-dependent, i.e. $\textbf{D}_{ij} = \textbf{D}_n$, with $n=1,2,3$. On the $\boldsymbol \alpha_n$ bond, the plane containing the Cr atoms and mediating ligands is a mirror plane of the bond; thus,  $\textbf{D}_n$ is perpendicular to this mirror plane:
\begin{equation} \label{dnn}
\textbf{D}_{n} = D_{n} \hat{\gamma}_n,
\end{equation}
  where $\hat{\gamma}_{1(2)} = \left(-\frac{1}{\sqrt{6}}, \pm \frac{1}{\sqrt{2}}, \frac{1}{\sqrt{3}}\right)$, and $\hat{\gamma}_{3} = \left(\sqrt{\frac{2}{3}}, 0, \frac{1}{\sqrt{3}}\right)$ are the unit vectors perpendicular to the mirror plane, depicted in Fig.~\ref{fig:structure}(b). 

The SOC also allows for a NN Kitaev interaction~\cite{aguilera2020},  which can be written as

\begin{equation} \label{kterm}
\mathcal{H}_{K}^{NN} = -\sum\limits_{\langle i,j \rangle} K_{ij} S_i^{\gamma_n} S_j^{\gamma_n},
\end{equation}
where $S_i^{\gamma_n}$ = $\textbf{S}_i \cdot \hat{\gamma}_n$ and $K_{ij} = K_n$. We can  combine Eqs.~\eqref{jterm}, \eqref{dterm}, and \eqref{kterm} by writing
\begin{equation} \label{comb1}
\mathcal{H}^{NN} = \mathcal{H}_{J}^{NN} + \mathcal{H}_{DM}^{NN} + \mathcal{H}_{K}^{NN} = \sum\limits_{\langle i,j \rangle} \textbf{S}_i^T \Lambda_{n} \textbf{S}_j,
\end{equation}
where
\begin{equation} \label{lambdamat}
\Lambda_{n} =
\begin{bmatrix}
-J_{n} & -D_{n}^z & D_{n}^y \\
D_{n}^z & -J_{n} & -D_{n}^x \\
-D_{n}^y & D_{n}^x & -J_{n}
\end{bmatrix} 
- K_n \hat{\gamma}_n \otimes \hat{\gamma}_n,
\end{equation}
is the NN interaction matrix, and $n$ is understood to index the $\langle i, j \rangle$ bond type.
The NNN Heisenberg and DM interactions can be included as
\begin{equation} \label{NNNjdterm}
\mathcal{H}^{NNN} = -\sum\limits_{\langle \langle i,j \rangle \rangle} J'_{ij} \textbf{S}_i \cdot \textbf{S}_j - \sum\limits_{\langle \langle i,j \rangle \rangle} \textbf{D}'_{ij} \cdot (\textbf{S}_i \times \textbf{S}_j)\,,
\end{equation}
where $\langle \langle .. \rangle \rangle$ denotes summation over next-to-nearest neighbors. Here, $J'_{ij}$ and $D'_{ij}$ are, respectively, the bond-dependent NNN Heisenberg and DM interaction strength.
There are three distinct NNN bonds on each of the two sublattices for a total of six possible NNN exchange parameters. For the sublattice $s=A,B$, the  bond along the hopping direction $\pm \boldsymbol 
\beta_n$, sketched in Fig.~\ref{fig:structure}(a), mediates a Heisenberg exchange $J'_{ij} = J'_{n,s}$ and a DM interaction $\textbf{D}'_{ij} = \pm \textbf{D}'_{n,s}$.
The lack of point-group symmetries provides no restriction on the NNN DM vectors according to Moriya's rules. Thus, the NNN DM vector $\textbf{D}'_{n,s}$ can be generally written in terms of the local bond geometry as
\begin{equation} \label{dnnn}
\begin{split}
\textbf{D}'_{n,s} &= \left(D'_{n,s} \sin{\tilde{\theta}_{n,s}}\right) R_z\left(\tilde{\phi}_{n,s}\right) \hat{\boldsymbol \beta}_{n} \\
&+ \tau_{s}\left(D'_{n,s} \cos{\tilde{\theta}_{n,s}}\right) \hat{\textbf{z}},
\end{split}
\end{equation}
where $ \hat{\boldsymbol \beta}_{n} = \boldsymbol \beta_{n}/| \boldsymbol \beta_{n} |$,  $R_z (\tilde{\phi}_{n,s})$ describes a right-handed rotation  by an angle $\tilde{\phi}_{n,s}$ about the $\hat{\mathbf{z}}$ axis and $\tau_{A(B)} = \pm 1$. The angles $\tilde{\theta}_{n,s}$ and $\tilde{\phi}_{n,s}$ are the spherical coordinates of $\textbf{D}'_{n,s}$ with azimuthal angle measured relative to the $\boldsymbol \beta_n$ bond on the $s$ sublattice; this geometry is shown in Fig.~\ref{fig:structure}(c). When the mediating halides are of the same type, the axis bisecting the bond vector through the mediating Cr is a two-way rotation axis, which constrains $\tilde{\phi}_{n,B} = 0$. 

We can rewrite Eq.~(\ref{NNNjdterm}) in a compact form as
\begin{equation} \label{NNNterms}
\mathcal{H}^{NNN} = \sum\limits_{\langle \langle i,j \rangle \rangle} \textbf{S}_i^T \Xi_{n,s} \textbf{S}_j,
\end{equation}
with
\begin{equation}\label{ximat}
\Xi_{n,s} =
\begin{bmatrix}
-J'_{n,s} & -D_{n,s}^{\prime z} & D_{n,s}^{\prime y} \\
D_{n,s}^{\prime z} & -J'_{n,s} & -D_{n,s}^{\prime x} \\
-D_{n,s}^{\prime y} & D_{n,s}^{\prime x} & -J'_{n,s}
\end{bmatrix}.
\end{equation}

Further, we include a single-ion anisotropy term, $\mathcal{H}_{A}$,  and a Zeeman interaction, $\mathcal{H}_{B}$,  due to a uniform external magnetic field $\mathbf{B}=B \hat{\mathbf{z}}$  as 
\begin{equation} \label{kbterm}
\mathcal{H}_A + \mathcal{H}_B = -A\sum\limits_i (S_i^z)^2 - g \mu_B B\sum\limits_i S_i^z.
\end{equation}
where $A>0$ parametrizes the strength of the easy-axis anisotropy~\cite{tartaglia2020}, $g$ is the g-factor  and $\mu_{B}$ is the Bohr magneton.

At each magnetic site, we can orient a spin-space Cartesian coordinate system such that the new $\hat{\mathbf{z}}$ axis locally lies along the classical orientation of the onsite spin operator $\tilde{\textbf{S}}_i$. The latter can be related to the spin operator $\textbf{S}_i $ in the global frame of reference via the transformation
\begin{equation} \label{spinvec}
\textbf{S}_i = R_i(\theta_i, \phi_i) \tilde{\textbf{S}}_i.
\end{equation}

Here, $R_i(\theta_i,\phi_i) = R_z(\phi_i)R_y(\theta_i)$, where $R_{z(y)} (\zeta)$ describes a right-handed rotation by an angle $\zeta$ about the global $\hat{\mathbf{z}}$ $(\hat{\mathbf{y}})$ axis,  and $\theta_i$ and $\phi_i$ are, respectively, the polar and azimuthal angles of the classical  orientation of the spin $\mathbf{S}_{i}$. Equations  \eqref{comb1}, \eqref{NNNterms} and \eqref{kbterm} can be combined into the full Hamiltonian in local coordinates as
\begin{equation} \label{HL}
\begin{split}
\mathcal{H} &= \sum\limits_{\langle i, j \rangle} \tilde{\textbf{S}}_i^T \tilde{\Lambda}_{n}\tilde{\textbf{S}}_j + \sum\limits_{\langle \langle i, j \rangle \rangle} \tilde{\textbf{S}}_i^T \tilde{\Xi}_{n,s} \tilde{\textbf{S}}_j \\
& - A\sum\limits_i \left( R_i \tilde{\textbf{S}}_i \right)^2_z -
	\mu_B B\sum\limits_i (R_i \tilde{\textbf{S}}_i)_z, \\
\end{split}
\end{equation}
where $\left( \cdot \right)_\mu$ is the $\mu$ component of a vector. Here, we have introduced the rotated interaction matrices  $\tilde{\Lambda}_n = R_i^T \Lambda_n R_j$ and $\tilde{\Xi}_{n,s} = R_i^T \Xi_{n,s} R_j$.

\subsection{BdG Hamiltonian}

Far below the magnetic ordering temperature $T_{c}$, i.e. for $T \ll T_c$, we can access the magnon spectrum by linearizing the Holstein-Primakoff transformation  \cite{holstein1940} in the local frame of reference, i.e. 
\begin{equation} \label{hptransform}
\begin{split}
\tilde{S}_i^+ &= \tilde{S}_i^x + i \tilde{S}_i^y = \sqrt{2S}\sqrt{1 - \frac{d_i^\dagger d_i}{2S}} d_i \approx \sqrt{2S} d_i, \\
\tilde{S}_i^z &= S - d_i^\dagger d_i\,,
\end{split}
\end{equation}
where $S$ is the classical spin (in units of $\hbar$) and $d_i$ ($d^{\dagger}_{i}$) the  magnon annihilation (creation) operator at the $i$th site, obeying the bosonic commutation relation ${[d_i, d_j^\dagger] = \delta_{ij}}$.  We plug Eq.~(\ref{hptransform}) into Eq.~\eqref{HL} and  truncate the  Hamiltonian beyond the quadratic terms in the Holstein-Primakoff boson operators since  interactions between magnons can be neglected in the temperature regime of interest. We group terms constant in magnon operators in the classical energy term $E_{Cl}(\{\theta_i, \phi_i\} \vert_i)$ \footnote{This is equivalent to regarding $\textbf{S}_i$ as classical spin vectors and equating the $\mathcal{H}$ with $E_{Cl}$.}. Minimization of $E_{Cl}$ with respect to $\{ \theta_i, \phi_i \} \vert_i$ gives the ground-state spin configuration.  Here, we focus on a ground state with two-sublattice translational symmetry, i.e.
\begin{equation}\label{spinAB}
\textbf{S}_i = S (\cos \phi_s \sin \theta_s, \sin \phi_s \sin \theta_s, \cos \theta_s),
\end{equation}
where $s=A,B$.
The classical energy then takes the form 
\begin{equation} \label{eclassical}
\begin{split}
&	E_{Cl}(\{\theta_i, \phi_i\} \vert_i)/N = E_{Cl}(\theta_A, \phi_A, \theta_B, \phi_B)/N \\
 	&= -g \mu_B B S (\cos \theta_A + \cos \theta_B) - A S^2 (\cos^2 \theta_A + \cos^2 \theta_B). \\
	&+ S \sum\limits_{n=1}^3 \tilde{\Lambda}_n^{zz},
\end{split}
\end{equation}
where $N$ is the total number of Cr atoms in the sample. Equation ~(\ref{eclassical}) can be minimized by gradient descent or Monte Carlo methods. 

In what follows, we relabel the operator $d_i$ as $a_i$ ($b_i$) on the $A$ $(B)$ sublattice. We can introduce the magnon operators in momentum space, i.e. $a_\textbf{k}$ and $b_\textbf{k}$,  by performing a Fourier transformation:

\begin{equation} \label{fouriertransform}
a_i = \sqrt{\frac{2}{N}}\sum\limits_{\textbf{k}} e^{i \textbf{k} \cdot \textbf{r}_i} a_\textbf{k}, \qquad b_i =\sqrt{\frac{2}{N}} \sum\limits_{\textbf{k}} e^{i \textbf{k} \cdot \textbf{r}_i} b_\textbf{k}\,,
\end{equation}
where $\mathbf{k}=(k_{x},k_{y})$ is the 2$d$ wavevector and the summation is taken over the first Brillouin zone.
Substituting Eq.~\eqref{fouriertransform} into the Hamiltonian~(\ref{HL}) yields

\begin{equation} \label{HBdG}
\mathcal{H} = \frac{1}{2}\sum\limits_{\textbf{k}} \psi_\textbf{k}^\dagger \mathcal{H}_{\text{BdG}}(\textbf{k}) \psi_\textbf{k},
\end{equation}
where $\psi_{\textbf{k}}^\dagger = 
\begin{bmatrix}
		a_\textbf{k}^\dagger, & b_\textbf{k}^\dagger, & a_{-\textbf{k}}, & 	b_{-\textbf{k}} \\
	\end{bmatrix}$ and
\begin{equation} \label{HBdGG}
\mathcal{H}_{\text{BdG}}(\textbf{k}) = 
\begin{bmatrix}
		h(\textbf{k}) & \Delta(\textbf{k}) \\
		\Delta^*(-\textbf{k}) & h^*(-\textbf{k}) \\
	\end{bmatrix},
\end{equation}
is a $4  \times 4$ Bogoliubov de Gennes  (BdG) Hamiltonian.
Here, $h(\textbf{k})$ and $\Delta(\textbf{k})$ are $2\times2$ matrices satisfying $h^\dagger(\textbf{k}) = h(\textbf{k})$ and $\Delta^T(\textbf{k}) = \Delta(-\textbf{k})$. Introducing

\begin{equation}\label{pmmats}
\begin{split}
\tilde{\Lambda}_n^\pm &= \tilde{\Lambda}_n^{xx} \pm \tilde{\Lambda}_n^{yy} + i (\tilde{\Lambda}_n^{yx} \mp \tilde{\Lambda}_n^{xy}), \\
\tilde{\Xi}_{n,s}^\pm &= \tilde{\Xi}_{n,s}^{xx} \pm \tilde{\Xi}_{n,s}^{yy} + i (\tilde{\Xi}_{n,s}^{yx} \mp \tilde{\Xi}_{n,s}^{xy}),
\end{split}
\end{equation}
the submatrices $h$ and $\Delta$ can be written explicitly as
\begin{equation}\label{hmatrix}
\begin{split}
	h_{11}(\textbf{k}) &= g \mu_B B \cos \theta_A + \frac{6 A S}{2} \cos^2 \theta_A - A S \\
	& - S\sum\limits_{n=1}^3[ \tilde{\Lambda}_n^{zz} + 2\tilde{\Xi}_{n,A}^{zz} - \text{Re} (\tilde{\Xi}_{n,A}^{+} e^{i \textbf{k} \cdot \boldsymbol \beta_n})], \\
	h_{22}(\textbf{k}) &= g \mu_B B \cos \theta_B + \frac{6 A S}{2} \cos^2 \theta_B - A S \\
	&  - S\sum\limits_{n=1}^3[ \tilde{\Lambda}_n^{zz} + 2\tilde{\Xi}_{n,B}^{zz} - \text{Re} (\tilde{\Xi}_{n,B}^{+} e^{i \textbf{k} \cdot \boldsymbol \beta_n})], \\
	h_{12}(\textbf{k}) &= \frac{S}{2}\sum\limits_{n=1}^3 \tilde{\Lambda}_n^+ e^{-i \textbf{k} \cdot \boldsymbol\alpha_n}, \; \; \; \; h_{21}(\textbf{k}) = h_{12}^*(\textbf{k}), \\
	\end{split}
\end{equation}

and

\begin{align}\label{dmatrix}
	\Delta_{11}(\textbf{k}) &= -AS \sin^2 \theta_A,  \; \; \; \; \Delta_{22}(\textbf{k}) = -AS \sin^2 \theta_B\,, \\
	\Delta_{12}(\textbf{k}) &= \frac{S}{2}\sum\limits_{n=1}^3 \tilde{\Lambda}_n^- e^{-i \textbf{k} \cdot \boldsymbol\alpha_n}, \; \; \; 
	\Delta_{21}(\textbf{k}) = \Delta_{12}(-\textbf{k})\,.
\end{align}

Since the system is bosonic, the Hamiltonian $\mathcal{H}_{\text{BdG}}(\textbf{k})$ must be diagonalized by a paraunitary BdG transformation~\cite{kohei2019,shindou2013,kondo2020h}. In other words, one should diagonalize the effective Hamiltonian 
\begin{align}
\tilde{\mathcal{H}}(\textbf{k}) = \Sigma_z \mathcal{H}_{\text{BdG}}(\textbf{k})\,, \; \; \; \Sigma_z = \sigma_z \otimes 1_{2 \times 2}\,,
\label{effH}
\end{align} 
where we have introduced the third Pauli matrix $\sigma_z$ and the $2 \times 2$ identity matrix $1_{2 \times 2}$.  We label the ${M=2}$ positive eigenvalues and associated eigenvectors of $\tilde{\mathcal{H}}(\textbf{k})$ as, respectively,  $\mathcal{E}_m (\textbf{k})$ and $\ket{m(\textbf{k})}$. The remaining $M$ states with negative eigenvalues $-\mathcal{E}_{m}(-\textbf{k})$ are an artifact of doubling the degrees of freedom and can be discarded.

\section{Topological magnons}
\subsection{Topological classification}

The topological classification of the Hermitian matrix $\mathcal{H}_{\text{BdG}}(\textbf{k})$ reduces to  the classification of the effective Hamiltonian $\tilde{\mathcal{H}}(\textbf{k})$, which is generally non-Hermitian~\cite{kohei2019, lieu2018}.  However, the  Hermiticity of the physical system guarantees that the effective matrix $\tilde{\mathcal{H}}(\textbf{k})$  has a built-in pseudo-Hermiticity symmetry, i.e.
\begin{equation} \label{PHH}
\eta^{-1} \tilde{\mathcal{H}}^{\dagger}(\textbf{k}) \eta = \tilde{\mathcal{H}}(\textbf{k}), \\  \quad \eta = \Sigma_z.
\end{equation}
\begin{table*}
	\centering
	\begin{tabularx}{\textwidth}{| Z | Z | Z | Z | Z | Z |}
		\hline\hline
		$S = 3/2$ & $g \mu_B B = 0.25$ & $A = 0.22$ & $J_1 = 1.2$ & $J_2 = 1.5$ & $J_3 = 1.8$ \\
		\hline
		$K_1 = 0.7$ & $K_2 = 0.5$ & $K_3 = 1.1$ & $D_1 = 0.2$ & $D_2 = 0.3$ & $D_3 = 0.6$ \\
		\hline
		$J'_{1,A} = 0.2$ & $J'_{2,A} = 0.4$ & $J'_{3,A} = 0.2$ & $J'_{1,B} = 0.1$ & $J'_{2,B} = 0.3$ & $J'_{3,B} = 0.4$ \\
		\hline
		$D'_{1,A} = 0.4$ & $D'_{2,A} = 0.2$ & $D'_{3,A} = 0.25$ & $D'_{1,B} = 0.5$ & $D'_{2,B} = 0.15$ & $D'_{3,B} = 0.05$ \\
		\hline
		$\tilde{\theta}_{1,A} = -0.17$ & $\tilde{\theta}_{2,A} = -0.07$ & $\tilde{\theta}_{3,A} = 0.22$ & $\tilde{\theta}_{1,B} = -0.37$ & $\tilde{\theta}_{2,B} = -0.47$ & $\tilde{\theta}_{3,B} = -0.57$ \\
		\hline
		$\tilde{\phi}_{1,A} = 0.3$ & $\tilde{\phi}_{2,A} = -0.8$ & $\tilde{\phi}_{3,A} = 0.2$ & $\tilde{\phi}_{1,B} = 0$ & $\tilde{\phi}_{2,B} = 0$ & $\tilde{\phi}_{3,B} = 0$ \\
		\hline\hline
	\end{tabularx}
\caption{Parameters used in the numerical diagonalization of $\tilde{\mathcal{H}}(\textbf{k})$~(\ref{effH}). All energy scales are in meV and angles are in radians.}
\label{table:values}
\end{table*}

Furthermore, the  Hamiltonian $\tilde{\mathcal{H}}(\textbf{k})$  obeys particle-hole symmetry (PHS), i.e.

\begin{equation} \label{PHS}
\mathcal{C} \tilde{\mathcal{H}}^T(\textbf{k}) \mathcal{C}^{-1} = -\tilde{\mathcal{H}}(-\textbf{k}), \\  \quad \mathcal{C} = \sigma_y \otimes 1_{2\times 2}\,.
\end{equation}

However, as discussed in detail by Refs.~\cite{kohei2019,lieu2018,lein2019}, for free bosons, particle-hole symmetry should be regarded as a built-in constraint of the Bogoliubov-de-Gennes Hamiltonian~\eqref{HBdGG}, rather than as a physical symmetry that can be selectively broken. Thus, the topological classification of $\tilde{\mathcal{H}}(\textbf{k})$ should effectively neglect Eq.~(\ref{PHS}).

When $\tilde{\Xi}^{xy}_{n,s} = \tilde{\Xi}^{yx}_{n,s} = 0$ and $\tilde{\Lambda}_n^{xy} = \tilde{\Lambda}_n^{yx} = 0$, the magnon Hamiltonian obeys  time-reversal symmetry, i.e. 
\begin{equation} \label{TRS}
\mathcal{T} \tilde{\mathcal{H}}^*(\textbf{k}) \mathcal{T}^{-1} = \tilde{\mathcal{H}}(-\textbf{k}), \\ \quad \mathcal{T} = 1_{4 \times 4}\,. 
\end{equation}

Generally, Eq.~(\ref{TRS}) holds in the absence of Kitaev or DM interactions, i.e. when $D'_{n,s} = D_{n} = K_{n} = 0$ for each $n$. 
In this case, the Hamiltonian belongs to the symmetry class $AI + \eta_{+}$~\cite{kohei2019}, which corresponds to a topologically trivial phase.

In the presence of finite Kitaev or  DM interaction, the relevant symmetry class is $A + \eta$~\cite{kohei2019}, which supports a topologically nontrivial phase characterized by a nonvanishing Chern number~\cite{shindou2013}. The (bosonic) Chern number of the $m$th band can be written as

\begin{equation} \label{chernc}
c_m = \frac{1}{2\pi}\int\limits_{BZ} d^2 \mathbf{k} \; \Omega_m^z (\textbf{k}), 
\end{equation}
where 
\begin{equation} \label{berrycurvature}
\boldsymbol\Omega_m (\textbf{k}) = \nabla_\mathbf{k} \times i \bra{m(\textbf{k})} \nabla_\mathbf{k} \ket{m(\textbf{k})},
\end{equation} 
is the Berry curvature on the $m$th band.

\begin{figure*}[ht]
	\centering
	\includegraphics[width=1.0\linewidth]{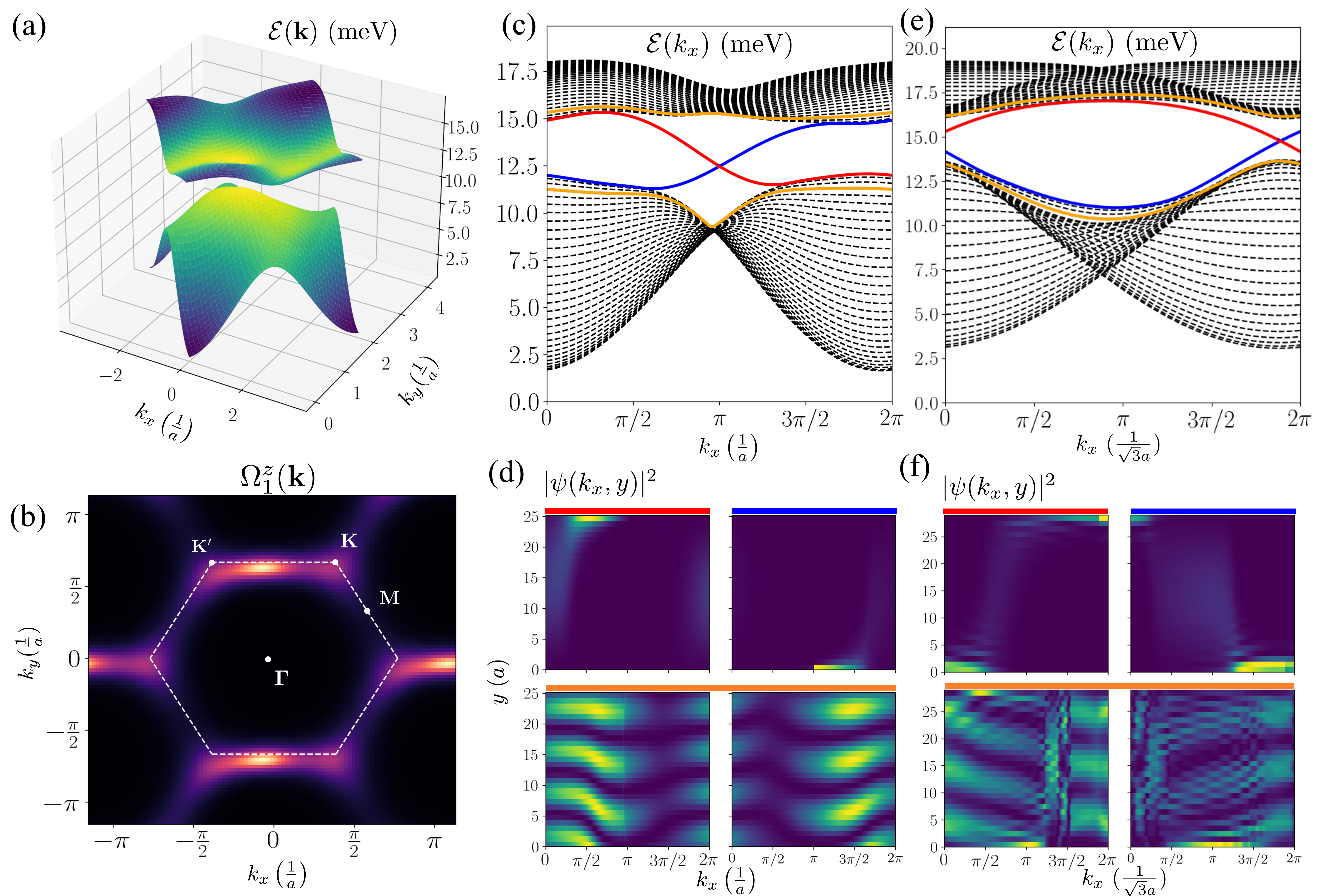}
	\caption{
	(a) Spin-wave dispersion.
	(b) The $z$-component $\Omega^{z}_{1}(\mathbf{k})$ of the Berry curvature~(\ref{berrycurvature}). The 1st Brillouin zone is indicated by a white hexagon. The local maxima of the Berry curvature are shifted off of the high symmetry points $\textbf{K}$ and $\textbf{K}'$ of the Brillouin zone due to  $C_{3}$ symmetry breaking. 
	(c-d) Exact diagonalization of Eq.~(\ref{HBdGG}) in a ribbon geometry with zigzag and armchair terminations, respectively, and 30 unit cells width. In both cases, the spectrum displays two topologically-protected edge states (blue and red line). Two bulk states are also highlighted in orange.
	(e-f) The eigenstates of the highlighted modes in (c-d) are shown. The edge states are exponentially confined to the top and bottom of the sample, whereas the bulk states are delocalized Bloch states. 
	}
	\label{fig:edgestates}
\end{figure*}

\begin{figure*}[ht!]
	\centering
	\includegraphics[width=1.0\textwidth]{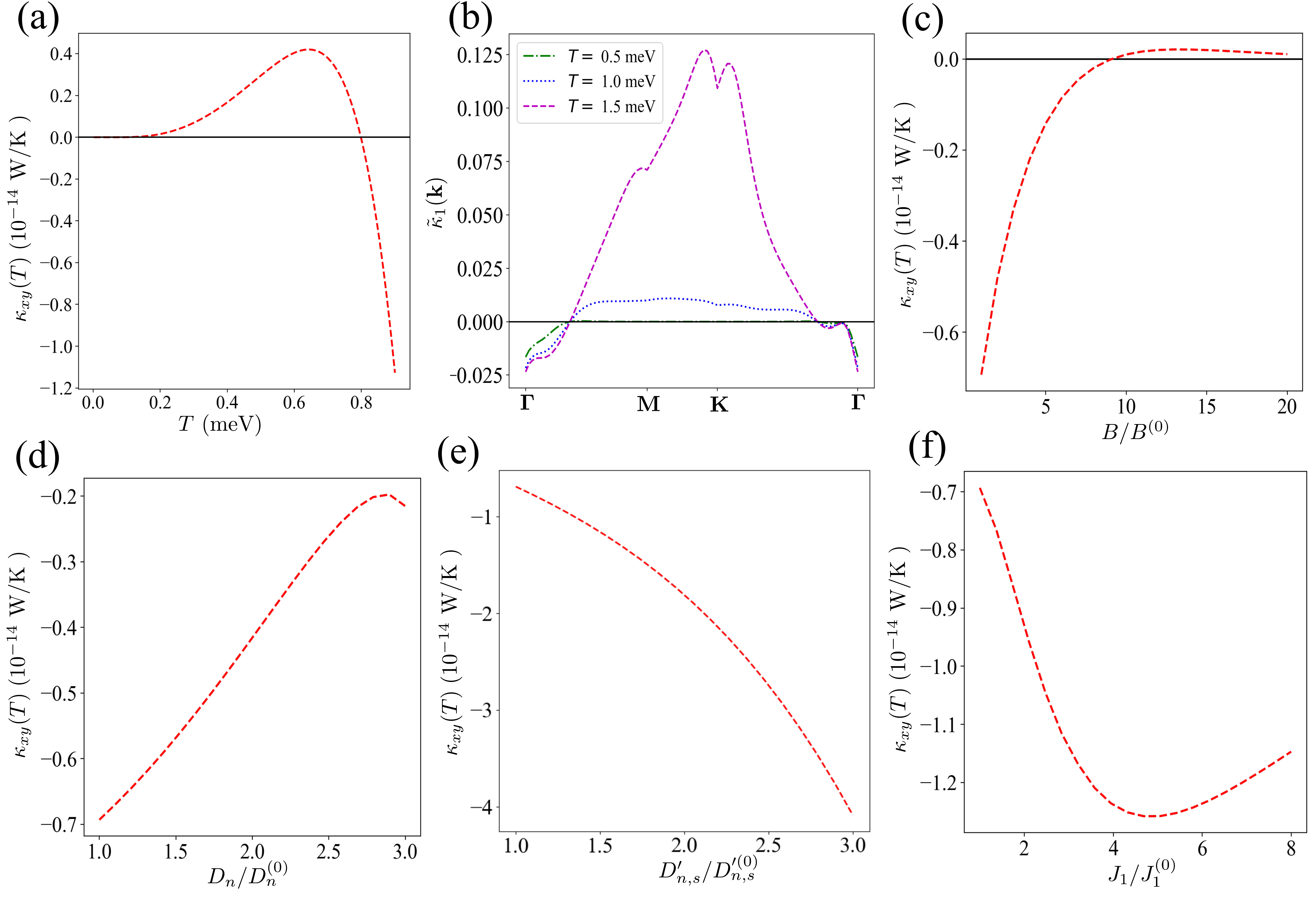}
	\caption{
	(a) Thermal Hall conductivity~(\ref{thermalhalleq}) as a function of temperature. 
	(b) The contribution of the lower band to the thermal Hall conductivity along a path of high symmetry in the BZ at various temperatures. In the subplots (c-f) the temperature is set at $T = 0.86$ meV and the $x$-axis of each subplot is a ratio of a parameter value to its initial value obtained in Table~\ref{table:values}, indicated by the superscript (0). The spins equilibrium positions are recalculated for each data point. Dependence of the thermal Hall conductivity~(\ref{thermalhalleq}) on the (c) magnetic field $B$;
	(d) NN DM magnitude $D_{n}$ ($D_{n}$ is increased for $n=1,2,3$, i.e. $D_{n}/D^{(0)}_{n}$ is equal for each bond);
	(e) NNN DM magnitude $D'_{n,s}$;  
	(f) NN Heisenberg exchange strength $J_{1}$.
	}
	\label{fig:thermalhall}
\end{figure*}

\subsection{Topological edge states}

Using the values in Table \ref{table:values}, the minimization of Eq.~\eqref{eclassical} by direct gradient descent yields the spin equilibrium positions $\theta_A \approx 0.41$, $\theta_B \approx 0.39$, $\phi_A \approx 0.18$, and $\phi_B \approx 0.18$. The bands acquire a nonzero Chern number, i.e. $c_{m}=\pm 1$ for $m=1(2)$.

We find that NN, NNN DM and Kitaev interactions can break time-reversal symmetry and open Chern-insulating gaps in the magnon spectrum. Figure \ref{fig:edgestates}(a) shows the gapped spectrum for the parameters of Table \ref{table:values}. Due to the lack of $C_{3}$ rotation symmetry, the Dirac nodes are not globally stable and the local maxima of the Berry curvature are shifted off the high symmetry point $\textbf{K}$ and $\textbf{K}'$, as shown in Fig.~\ref{fig:edgestates}(b). By varying the anisotropy of our parameters, we find that the two Dirac nodes can meet up and annihilate at the $\textbf{M}$ point.

The open boundary condition spectrum that results from exact diagonalization of Eq.~(\ref{effH}) in a ribbon geometry with zig-zag and armchair edges are presented in Fig.~\ref{fig:edgestates}(c-d). Two topologically-protected dispersive magnon  modes, localized at the edges of the ribbon (see Fig.~\ref{fig:edgestates}(e-f)), emerge as consequence of the topologically nontrivial character of the magnon bands.
\setlength{\extrarowheight}{2pt}

\subsection{Thermal Hall effect}

It is well known that a temperature gradient can induce a magnon transverse heat current in systems with topologically nontrivial magnon bands~\cite{katsura2010,onose2010,laurell2018,owerre2016,matsumoto2014,mook2014,moulsdale2019}. The (intrinsic) magnon thermal Hall conductivity can be calculated as~\cite{murakami2017}
\begin{equation} \label{thermalhalleq} 
\kappa_{xy}(T) = - \frac{T}{4 \pi^2} \sum\limits_{m = 1}^2 \int\limits_{BZ} d^2 \mathbf{k} \; \Omega_m^z (\textbf{k}) c_2 \left[ g_T(\mathcal{E}_m (\textbf{k})) \right],
\end{equation}
where $k_{B} = \hbar = 1$,  $g_T(x) = (e^{x/T}-1)^{-1}$ is the Bose-Einstein distribution function and 
\begin{equation} \label{c2}
c_2(x) = (1 + x) \left[ \log{\left( \frac{1+x}{x} \right) } \right]^2 - \left( \log{x} \right)^2 - 2 \text{Li}_2 (-x).
\end{equation}

Here, $\text{Li}_s (z)$ is the polylogarithm of order $s$ and argument $z$.
Figure~\ref{fig:thermalhall}(a) shows that, at low temperature, $\kappa_{xy}(T)$ displays a surprising change of sign. The sign change can be understood by rewriting Eq.~\eqref{thermalhalleq} as

\begin{equation}
\begin{split}
\kappa_{xy}(T) &= - \frac{T}{4 \pi^2} \sum\limits_{m = 1}^2 \int\limits_{BZ} d^2 \mathbf{k} \; \tilde{\kappa}_{m}(\textbf{k}), \\
\tilde{\kappa}_{m}(\textbf{k}) &= \Omega_m^z(\textbf{k})c_2\left[g_T(\mathcal{E}_n(\textbf{k}))\right].
\end{split}
\end{equation}

Here, $\tilde{\kappa}_m(\textbf{k})$ is proportional to the contribution to $\kappa_{xy}(T)$ from the $m$th band at the momentum $\textbf{k}$. Since $c_2$ is positive and monotonically increasing, the sign of $\tilde{\kappa}_m(\textbf{k})$ depends only on $\Omega_m^z(\textbf{k})$. 
For the lower magnon band,  the Berry curvature $\Omega_1^z(\textbf{k})$ has negative sign in the neighborhood of the $\boldsymbol \Gamma$ point, while it is positive around the gap-closing points near $\textbf{K}$ and $\textbf{K}'$. At lower temperatures, only states in the lower band in the vicinity of the $\boldsymbol \Gamma$ point are 
populated. The factor of $c_2[g_T(\mathcal{E}_n(\textbf{k}))]$ suppresses finite contribution to $\tilde{\kappa}_1(\textbf{k})$ at reciprocal lattice points except those close to $
\boldsymbol \Gamma$.  As $T$ increases, the states at the gap-closing points near $\textbf{K}$ and $\textbf{K}'$ become populated and, due to their large negative Berry curvature, come to dominate $\tilde{\kappa}_1(\textbf{k})$. This leads to the sign change of the thermal Hall conductivity $\kappa_{xy}(T)$  at $T \approx 0.7$ meV, shown in Fig.~\ref{fig:thermalhall}(b).

Another sign change in the thermal Hall conductivity $\kappa_{xy}$ occurs when the magnitude of the magnetic field is increased, as depicted in Fig.~\ref{fig:thermalhall}(c).  Increasing the magnetic field yields to an overall shift of the bands to higher energies. As a result, states that once populated the region near $\textbf{K}$ become energetically unfavorable while states near $\boldsymbol \Gamma$ remain populated, thus causing the sign of $\kappa_{xy}$ to change. 

The influence of the NNN and NN DM interaction on the thermal Hall flow is depicted, respectively, in Fig.~\ref{fig:thermalhall}(d) and Fig.~\ref{fig:thermalhall}(e). The NN (NNN) DM interaction change both the matrix elements of $\Lambda_n$ ($\Xi_{n,s}$) as well as the ground state configuration, which in turn modifies the overall structure of $\boldsymbol \Omega_m$ and $\mathcal{E}_m$ . The result is that $\tilde{\kappa}_1$ near $\boldsymbol \Gamma$, which is the primary contribution to $\kappa_{xy}$, increases with $D'_{n,s}/D^{\prime (0)}_{n,s}$ and decreases with $D_{n}/D^{(0)}_{n}$. Increasing either DM magnitude further causes the ground state to leave the uniform regime and our earlier assumption of two-sublattice translational symmetry breaks down.

In Fig.~\ref{fig:thermalhall}(f), the NN Heisenberg exchange along $\boldsymbol \alpha_1$ is increased. Initially, this leads to $\kappa_{xy}$ increasing, but around $J_1/J_1^{(0)} \approx 3$, the anisotropy becomes high enough to push the Dirac nodes together at $\textbf{M}$, where they annihilate, and the system enters a topologically trivial phase.

\begin{figure}[t!]
	\includegraphics[width=0.5\textwidth]{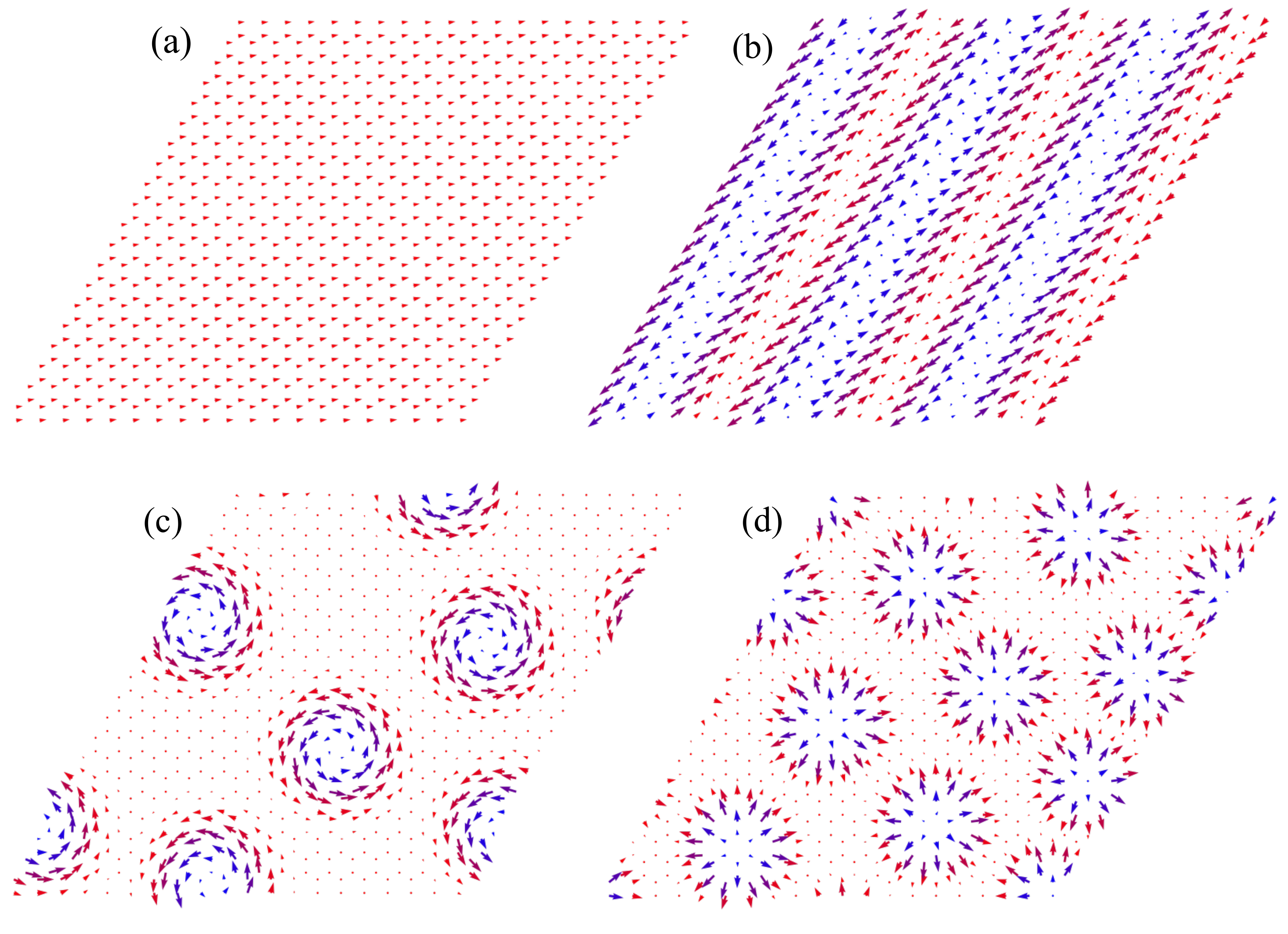}
	\caption{
	Ground state spin textures obtained by MCMC. Each plot shows the classical spin moments projected onto the $xy$ plane, where blue lines indicate a positive $z$-component and red lines indicate a negative $z$-component.
	(a) The ground state for the values given in Table~\ref{table:values}.
	(b) A spin cycloid.
	(c-d) Bloch and N\'{e}el skyrmions, respectively. }
	\label{fig:montecarlo}
\end{figure}

\section{Monte Carlo simulations}

Throughout our discussion, we have focused on a ground state with a two-sublattice translational symmetry and we have shown that the symmetry-breaking interactions, i.e., NN and NNN DM and Kitaev, can give rise to topologically nontrivial spin-wave bands. In this last section, we show that changing the strength and/or the anisotropy of the symmetry-breaking spin interactions can yield  spin textures that have a nontrivial real-space topology.
The large parameter space allows for a wide variety of noncollinear ground states that can be accessed by Markov-Chain Monte Carlo (MCMC)~\cite{xu2020, liang2020}, which we  have used to verify that the values given in Table~\ref{table:values}  correspond to a two-sublattice ground state. 

Taking a 20$\times$20 lattice subject to periodic boundary conditions, we perform annealed Metropolis MCMC followed by gradient descent, guaranteeing that the solution is at least a local minima (metastable state), if not the true ground state. In Fig.~\ref{fig:montecarlo}(a), we show the ground state using values obtained in Table~\ref{table:values} has a two-sublattice periodicity; the polar and azimuthal angles of the spin moments agree to within 1\% of those obtained by gradient descent. 
In the remaining figures, we explore other parameter regimes. Fig~\ref{fig:montecarlo}(b) shows a spin cycloid, while Fig~\ref{fig:montecarlo}(c-d) show Bloch and N\'{e}el skyrmions, which emerge when there is a strong enough NNN or NN DMI, respectively \cite{yu2010, heinze2011, rler2006, kzsmrki2015}.

\section{Conclusions} 
In this work, we have constructed a model for a CrClBrI monolayer, though an appropriate choice of parameters reduces our model to a generic two-sublattice translationally symmetric CrCl$_{3-x-y}$Br$_x$I$_y$ monolayer. Focusing on a linear spin-wave regime and on a ground state with a sublattice unit cell, we have shown that (both NN and NNN) DMI and the Kitaev interactions can drive the system into a magnon Chern insulating phase.  The topologically-protected magnon edge states associated with nonvanishing Chern numbers yield a thermal Hall effect. We find that the sign of the thermal Hall conductivity can be controlled by tuning temperature and external magnetic fields. 

Finally, we show that our spin model can support a variety of ground states depending on the choice of parameters, including magnetic topological defects. Chromium trihalides have been also proposed as possible hosts of quantum spin liquids (QSL)  \cite{xu2018,xu2020:kitaev}. However,  the experimental results of Tartaglia \textit{et al.} \cite{tartaglia2020} show that CrClBrI has a frustration index of $f \sim 2$, which suggests that the magnetic interactions are not sufficiently frustrated to support a QSL ground state. It is also worth noting that, while it may be possible in principle for the model presented in the present work to support a QSL ground state, our Monte-Carlo simulations show that -- for the parameters considered in this analysis -- the ground state spin arrangement is not frustrated.

We hope that our results will stimulate  systematic  \textit{ab initio} and experimental investigations of the coupling strengths introduced in our model.

\section{Acknowledgments}
The authors thank F. Tafti for insightful discussions.

\bibliographystyle{prsty}
\bibliography{bibliography}

\begin{thebibliography}{10}

\bibitem{burch2018}
K.~S. Burch, D. Mandrus, and J.-G. Park, Nature {\bf 563},  47  (2018).

\bibitem{gong2017}
C. Gong, L. Li, Z. Li, H. Ji, A. Stern, Y. Xia, T. Cao, W. Bao, C. Wang, Y.
  Wang, Z.~Q. Qiu, R.~J. Cava, S.~G. Louie, J. Xia, and X. Zhang, Nature {\bf
  546},  265  (2017).

\bibitem{huang2017}
B. Huang, G. Clark, E. Navarro-Moratalla, D.~R. Klein, R. Cheng, K.~L. Seyler,
  D. Zhong, E. Schmidgall, M.~A. McGuire, D.~H. Cobden, W. Yao, D. Xiao, P.
  Jarillo-Herrero, and X. Xu, Nature {\bf 546},  270  (2017).

\bibitem{park2016}
J.-G. Park, Journal of Physics: Condensed Matter {\bf 28},  301001  (2016).

\bibitem{mermin1966}
H.~W. Nathaniel D.~Mermin, Phys. Rev. Lett. {\bf 17},  1133  (1966).

\bibitem{hohenberg1967}
P.~C. Hohenberg, Phys. Rev. {\bf 158},  383  (1967).

\bibitem{lee2016}
J.-U. Lee, S. Lee, J.~H. Ryoo, S. Kang, T.~Y. Kim, P. Kim, C.-H. Park, J.-G.
  Park, and H. Cheong, Nano Letters {\bf 16},  7433  (2016).

\bibitem{wang2016}
X. Wang, K. Du, Y.~Y.~F. Liu, P. Hu, J. Zhang, Q. Zhang, M.~H.~S. Owen, X. Lu,
  C.~K. Gan, P. Sengupta, C. Kloc, and Q. Xiong, 2D Materials {\bf 3},  031009
  (2016).

\bibitem{bonilla2018}
M. Bonilla, S. Kolekar, Y. Ma, H.~C. Diaz, V. Kalappattil, R. Das, T. Eggers,
  H.~R. Gutierrez, M.-H. Phan, and M. Batzill, Nature Nanotechnology {\bf 13},
  289  (2018).

\bibitem{ohara2018}
D.~J. O'Hara, T. Zhu, A.~H. Trout, A.~S. Ahmed, Y.~K. Luo, C.~H. Lee, M.~R.
  Brenner, S. Rajan, J.~A. Gupta, D.~W. McComb, and R.~K. Kawakami, Nano
  Letters {\bf 18},  3125  (2018).

\bibitem{berezinskii1971}
V.~L. Berezinskii, Soviet Phys. JETP {\bf 32},  493  (1971).

\bibitem{kosterlitz1973}
J.~M. Kosterlitz and D.~J. Thouless, J. Phys. C. {\bf 6},  1181  (1973).

\bibitem{kosterlitz1974}
J.~M. Kosterlitz, J. Phys. C. {\bf 7},  1046  (1974).

\bibitem{kim2021}
S.~K. Kim and S.~B. Chung, SciPost Phys. {\bf 10},  68  (2021).

\bibitem{troncoso2020}
R.~E. Troncoso, A. Brataas, and A. Sudb\o{}, Phys. Rev. Lett. {\bf 125},
  237204  (2020).

\bibitem{pershoguba2018}
S.~S. Pershoguba, S. Banerjee, J.~C. Lashley, J. Park, H. \AA{}gren, G. Aeppli,
  and A.~V. Balatsky, Phys. Rev. X {\bf 8},  011010  (2018).

\bibitem{kitaev2006}
A. Kitaev, Annals of Physics {\bf 321},  2  (2006).

\bibitem{xu2018}
C. Xu, J. Feng, H. Xiang, and L. Bellaiche, npj Computational Materials {\bf
  4},  57  (2018).

\bibitem{lee2020}
I. Lee, F.~G. Utermohlen, D. Weber, K. Hwang, C. Zhang, J. van Tol, J.~E.
  Goldberger, N. Trivedi, and P.~C. Hammel, Phys. Rev. Lett. {\bf 124},  017201
   (2020).

\bibitem{xu2020:kitaev}
C. Xu, J. Feng, M. Kawamura, Y. Yamaji, Y. Nahas, S. Prokhorenko, Y. Qi, H.
  Xiang, and L. Bellaiche, Phys. Rev. Lett. {\bf 124},  087205  (2020).

\bibitem{chen2018}
L. Chen, J.-H. Chung, B. Gao, T. Chen, M.~B. Stone, A.~I. Kolesnikov, Q. Huang,
  and P. Dai, Phys. Rev. X {\bf 8},  041028  (2018).

\bibitem{kim2016}
S.~K. Kim, H. Ochoa, R. Zarzuela, and Y. Tserkovnyak, Phys. Rev. Lett. {\bf
  117},  227201  (2016).

\bibitem{ruckriegel2018}
A. R\"uckriegel, A. Brataas, and R.~A. Duine, Phys. Rev. B {\bf 97},  081106
  (2018).

\bibitem{wang2018}
Z. Wang, T. Zhang, M. Ding, B. Dong, Y. Li, M. Chen, X. Li, J. Huang, H. Wang,
  X. Zhao, Y. Li, D. Li, C. Jia, L. Sun, H. Guo, Y. Ye, D. Sun, Y. Chen, T.
  Yang, J. Zhang, S. Ono, Z. Han, and Z. Zhang, Nature Nanotechnology {\bf 13},
   554  (2018).

\bibitem{behera2019}
A.~K. Behera, S. Chowdhury, and S.~R. Das, Applied Physics Letters {\bf 114},
  232402  (2019).

\bibitem{lu2017}
A.-Y. Lu, H. Zhu, J. Xiao, C.-P. Chuu, Y. Han, M.-H. Chiu, C.-C. Cheng, C.-W.
  Yang, K.-H. Wei, Y. Yang, Y. Wang, D. Sokaras, D. Nordlund, P. Yang, D.~A.
  Muller, M.-Y. Chou, X. Zhang, and L.-J. Li, Nature Nanotechnology {\bf 12},
  744  (2017).

\bibitem{liu2018}
J. Liu, M. Shi, P. Mo, and J. Lu, {AIP} Advances {\bf 8},  055316  (2018).

\bibitem{zhong2017}
D. Zhong, K.~L. Seyler, X. Linpeng, R. Cheng, N. Sivadas, B. Huang, E.
  Schmidgall, T. Taniguchi, K. Watanabe, M.~A. McGuire, W. Yao, D. Xiao,
  K.-M.~C. Fu, and X. Xu, Science Advances {\bf 3},  e1603113  (2017).

\bibitem{hellman2017}
F. Hellman, A. Hoffmann, Y. Tserkovnyak, G.~S.~D. Beach, E.~E. Fullerton, C.
  Leighton, A.~H. MacDonald, D.~C. Ralph, D.~A. Arena, H.~A. D\"urr, P.
  Fischer, J. Grollier, J.~P. Heremans, T. Jungwirth, A.~V. Kimel, B. Koopmans,
  I.~N. Krivorotov, S.~J. May, A.~K. Petford-Long, J.~M. Rondinelli, N.
  Samarth, I.~K. Schuller, A.~N. Slavin, M.~D. Stiles, O. Tchernyshyov, A.
  Thiaville, and B.~L. Zink, Rev. Mod. Phys. {\bf 89},  025006  (2017).

\bibitem{abramchuk2018}
M. Abramchuk, S. Jaszewski, K.~R. Metz, G.~B. Osterhoudt, Y. Wang, K.~S. Burch,
  and F. Tafti, Advanced Materials {\bf 30},  1801325  (2018).

\bibitem{kondo2020d}
H. Kondo and Y. Akagi, arXiv:2012.02034  (2020).

\bibitem{tartaglia2020}
T.~A. Tartaglia, J.~N. Tang, J.~L. Lado, F. Bahrami, M. Abramchuk, G.~T.
  McCandless, M.~C. Doyle, Y. Ran, J.~Y. Chan, and F. Tafti, Science Advances
  {\bf 6},  30  (2020).

\bibitem{katsura2010}
H. Katsura, N. Nagaosa, and P.~A. Lee, Phys. Rev. Lett. {\bf 104},  066403
  (2010).

\bibitem{matsumoto2011}
R. Matsumoto and S. Murakami, Phys. Rev. B {\bf 84},  184406  (2011).

\bibitem{murakami2017}
S. Murakami and A. Okamoto, Journal of the Physical Society of Japan {\bf 86},
  011010  (2017).

\bibitem{onose2010}
Y. Onose, T. Ideue, H. Katsura, Y. Shiomi, N. Nagaosa, and Y. Tokura, Science
  {\bf 329},  297  (2010).

\bibitem{moriya1960}
T. Moriya, Phys. Rev. {\bf 120},  91  (1960).

\bibitem{aguilera2020}
E. Aguilera, R. Jaeschke-Ubiergo, N. Vidal-Silva, L.~E. F.~F. Torres, and A.~S.
  Nunez, Phys. Rev. B {\bf 102},  024409  (2020).

\bibitem{holstein1940}
T. Holstein and H. Primakoff, Phys. Rev. {\bf 58},  1098  (1940).

\bibitem{Note1}
This is equivalent to regarding $\protect \textbf {S}_i$ as classical spin
  vectors and equating the $\protect \mathcal {H}$ with $E_{Cl}$.

\bibitem{kohei2019}
K. Kawabata, K. Shiozaki, M. Ueda, and M. Sato, Phys. Rev. X {\bf 9},  041015
  (2019).

\bibitem{shindou2013}
R. Shindou, R. Matsumoto, S. Murakami, and J.-i. Ohe, Phys. Rev. B {\bf 87},
  174427  (2013).

\bibitem{kondo2020h}
H. Kondo, Y. Akagi, and H. Katsura, Progress of Theoretical and Experimental
  Physics {\bf 2020},    (2020).

\bibitem{lieu2018}
S. Lieu, Phys. Rev. B {\bf 97},  045106  (2018).

\bibitem{lein2019}
M. Lein and K. Sato, Phys. Rev. B {\bf 100},  075414  (2019).

\bibitem{laurell2018}
P. Laurell and G.~A. Fiete, Physical Review B {\bf 98},  094419  (2018).

\bibitem{owerre2016}
S.~A. Owerre, Journal of Applied Physics {\bf 120},  043903  (2016).

\bibitem{matsumoto2014}
R. Matsumoto, R. Shindou, and S. Murakami, Physical Review B {\bf 89},  054420
  (2014).

\bibitem{mook2014}
A. Mook, J. Henk, and I. Mertig, Phys. Rev. B {\bf 89},  134409  (2014).

\bibitem{moulsdale2019}
C. Moulsdale, P.~A. Pantale{\'{o}}n, R. Carrillo-Bastos, and Y. Xian, Physical
  Review B {\bf 99},  214424  (2019).

\bibitem{xu2020}
C. Xu, J. Feng, S. Prokhorenko, Y. Nahas, H. Xiang, and L. Bellaiche, Phys.
  Rev. B {\bf 101},  060404  (2020).

\bibitem{liang2020}
J. Liang, W. Wang, H. Du, A. Hallal, K. Garcia, M. Chshiev, A. Fert, and H.
  Yang, Phys. Rev. B {\bf 101},  184401  (2020).

\bibitem{yu2010}
X.~Z. Yu, Y. Onose, N. Kanazawa, J.~H. Park, J.~H. Han, Y. Matsui, N. Nagaosa,
  and Y. Tokura, Nature {\bf 465},  901  (2010).

\bibitem{heinze2011}
S. Heinze, K. von Bergmann, M. Menzel, J. Brede, A. Kubetzka, R. Wiesendanger,
  G. Bihlmayer, and S. Bl\"{u}gel, Nature Physics {\bf 7},  713  (2011).

\bibitem{rler2006}
U.~K. R\"{o}{\ss}ler, A.~N. Bogdanov, and C. Pfleiderer, Nature {\bf 442},  797
   (2006).

\bibitem{kzsmrki2015}
I. K{\'{e}}zsm{\'{a}}rki, S. Bord{\'{a}}cs, P. Milde, E. Neuber, L.~M. Eng,
  J.~S. White, H.~M. R{\o}nnow, C.~D. Dewhurst, M. Mochizuki, K. Yanai, H.
  Nakamura, D. Ehlers, V. Tsurkan, and A. Loidl, Nature Materials {\bf 14},
  1116  (2015).

\end{thebibliography}

\end{document}